# Vortex matter as a magnetically programmable metasurface for tuning near-field radiative heat transfer

Siddharth Sangi[1], Pramod Reddy[2], and Shanhui Fan[3,†]

[1]*Huron High School, Ann Arbor, MI 48105, USA.*

[2]*Department of Mechanical Engineering, University of Michigan, Ann Arbor, MI 48109, USA.*

[3]*Department of Electrical Engineering, Ginzton Laboratory, Stanford University, Stanford, CA 94305, USA.*

ABSTRACT. Near-field radiative heat transfer (NFRHT) is of great interest for energy conversion and thermal management. Recent work on NFRHT has shown that metasurfaces can enhance NFRHT and reshape its spectral characteristics. Here, we show that the vortex lattice of a type-II superconductor represents a self-assembled metasurface whose lattice dimensions, in contrast to most other metasurfaces, can be tuned by magnetic fields, allowing NFRHT to be varied by orders of magnitude. Specifically, we show that for two Nb surfaces at 4.2 K and separated by a 100 nm gap, NFRHT can be magnetically tuned by a factor >200. Further, we show that the tuning range grows upon cooling, exceeding $10^5$ at 2 K, which is far beyond the order-unity modulation of many proposed schemes for tuning NFRHT. Our results establish vortex matter as a magnetically programmable thermal metasurface and open new opportunities for tuning the flow of heat at the nanoscale.

Near-field radiative heat transfer (NFRHT) arises when two bodies are separated by a vacuum gap smaller than Wien's wavelength [1,2] and is of great current interest [3–5] due to the novel transport phenomena that arise at the nanoscale and due to potential applications of NFRHT for energy conversion [6]. Experimental work [7–10] has unambiguously established that, in the near field, evanescent and surface modes enhance heat transfer rates over the blackbody limit by orders of magnitude. Recent theoretical work has extensively explored the possibility of tuning NFRHT via both nanostructuring and external stimuli. For example, theoretical work on nanostructured surfaces [11–14] has shown that introducing periodic variations in the dielectric properties of a surface can tune both the spectral characteristics and the magnitude of NFRHT. Further, very recent experimental work [15] has demonstrated that NFRHT can be strongly enhanced by metamaterials. However, the geometry of all these metasurfaces and metastructures is fixed during fabrication and post-fabrication control of the geometry has been limited to mechanical deformations of surfaces/structures via strain, twist, or lateral displacement [16–18], which enable modulation of NFRHT by factors of order unity.

In addition to employing metasurfaces for enhancing NFRHT, past work has also explored approaches for actively controlling NFRHT through temperature-induced metal-insulator [19–21] and superconducting [22–25] phase transitions, electrostatic gating [26,27], ferroelectric switching [28], and many-body effects [29,30]. Magneto-optic approaches [31,32], on the other hand, utilize multi-tesla fields (~6 T) to shift the cyclotron resonances and induce hyperbolic modes in magneto-optical semiconductors to suppress NFRHT by up to a factor of 7 in InSb [31]. In all these cases an external stimulus perturbs a fixed structure, and the resulting modulation of radiative heat transfer coefficient is typically of order unity, reaching at most a factor of a few in experiment and about seven in theory [31]. The one exception that could offer larger modulation depths is the superconducting transition [22,23], which switches the heat transfer on or off as the superconductor is warmed or cooled through $T_c$. However, cycling through $T_c$ requires changing the switch's temperature, which is undesirable for thermal switching applications. Instead, a thermal switch should operate at a fixed temperature with an external control parameter.

In this Letter, we show that the vortex lattice of a type-II superconductor acts as a self-assembled metasurface whose structure can be tuned continuously by an external magnetic field, at fixed temperature, to modulate NFRHT by more than two orders of magnitude. The key idea is illustrated in Fig. 1(a), which schematically depicts two cuboid-shaped macroscopic type-II superconductors (oxygen-doped Nb in this work), separated by a gap, and subject to a magnetic field $H$ applied along the $z$-axis, normal to the two parallel facing surfaces. We assume that the superconductors are sufficiently long along $z$ that demagnetization effects can be neglected. When the applied field $H$ exceeds the lower critical field ($H_{c1}$), quantized flux lines enter the superconductor and self-assemble into a triangular lattice of vortices whose spacing is set solely by the flux density [33]. Each vortex carries an approximately normal core embedded in a superconducting matrix, so the mixed state constitutes a self-assembled composite of lossy inclusions in a low-loss matrix. In contrast to nanofabricated metasurfaces, whose dimensions are fixed, and to active schemes, where an external field perturbs a fixed structure, the vortex lattice constant ($a_0$ in Fig. 1(a)) can be actively tuned by the applied field $H$. We note that near-field heat exchange between a nanoparticle and a superconductor has recently been proposed as a probe of the vortex structure [34]. Here, by contrast, the vortex lattice serves as a reconfigurable metasurface that controls NFRHT.

We use fluctuational electrodynamics [2,35] and rigorous coupled-wave analysis (RCWA) [36–38] to quantify the tunability of NFRHT in this system. We consider the system shown in Fig. 1(a), which consists of two Nb cuboids separated by a vacuum gap of $d$ = 100 nm and held at $T$ = 4.2 K. Further, we assume that the Nb contains ~0.96 at.% interstitial oxygen, giving a low-temperature (impurity-limited) residual resistivity of $\rho_0 = 5$ μΩ cm, a mean free path $\ell \approx$ 7.5 nm, critical temperature $T_c = 8.5$ K, Ginzburg-Landau (GL) parameter $\kappa \approx 4.3$, lower critical field $\mu_0 H_{c1} = 45$ mT and $\mu_0 H_{c2} = 0.79$ T (see Ref. [39] and SM, Sec. S1). Since $\ell \ll \xi_0$ (39 nm, clean-limit coherence length), λ and $d$ ($\ell/\xi_0 \approx 0.2$, $\ell/\lambda \approx 0.09$, $\ell/d \approx 0.08$, see SM, Sec. S1), Nb is in the dirty limit and its response is local, so the use of fluctuational electrodynamics with a local dielectric function is appropriate [40].

To analyze NFRHT in this configuration, it is essential to know the local dielectric function of Nb, $\varepsilon_{Nb}(\mathbf{r}, \omega)$, as a function of the applied $H$ field, where $\omega$ is the angular frequency and $\mathbf{r}$ is the position vector in the $xy$ plane. Since the vortex lattice is assumed to be invariant along the $z$-axis, $\varepsilon_{Nb}$ varies only within the $xy$ plane. Further, $H$ and the average flux density $\bar{B}$ are collinear with the $z$-axis, we therefore work with their $z$-components throughout. In order to obtain $\varepsilon_{Nb}(\mathbf{r}, \omega)$ we begin by first evaluating the mean flux density $\bar{B}(H)$, the vortex-lattice spacing $a_0 = (2\Phi_0/\sqrt{3}\ \bar{B})^{1/2}$ ($\Phi_0 = h/2e$), and the position-dependent order parameter $f(\mathbf{r})$ of the triangular lattice (see Fig. 1), using the Ginzburg–Landau (GL) approach for vortex lattices from Ref. [41] (see Supplemental Material [SM], Sec. S2 for details). The computed $f(\mathbf{r})$ is shown in Fig. 1(b) for $\mu_0 H$ = 150 mT (left panel) and 500 mT (right panel), for which the lattice constant ($a_0$) ≈ 134 nm and 70 nm, respectively — comparable to the gap size $d$.

Next, we obtain $\varepsilon_{Nb}(\mathbf{r}, \omega)$ from $f(\mathbf{r})$, by adapting the approach in Refs. [42,43], in which the optical conductivity $\sigma(\omega)$ is related to the pair potential $\bar{\Delta}$, a pair-breaking scattering rate $\Gamma(\mathbf{r})$, and a pair-conserving scattering rate $\Gamma_s$ (11.3 meV, see SM, Sec. S3). Specifically, we evaluate the local optical conductivity $\sigma(\mathbf{r},\omega)$ by obtaining the local pair potential, $\bar{\Delta}(\mathbf{r}) = f(\mathbf{r})\Delta_0$ [33] and the pair-breaking rate $\Gamma(\mathbf{r}) = \Gamma_0 + \Gamma_i(\mathbf{r})$. Here, $\Delta_0 = 1.3$ meV is the superconducting gap at 4.2 K (see Supplemental Material [SM], Secs. S1 and S3 for details), while $\Gamma_0 \approx 0.01$ meV is the zero-field rate inferred for Nb [44] (our 4.2 K results are insensitive to $\Gamma_0$ as described in Supplemental Material [SM], Sec. S3). The field-induced rate $\Gamma_i$, on the other hand, follows from the relation $\overline{\Delta} = f(\mathbf{r})\Delta_0 = [\Delta_0(\Delta_0 - 2\Gamma_i(\mathbf{r}))]^{1/2}$ [42,43] and the pair-breaking rate $\Gamma(\mathbf{r})$ can be written as:

$$\Gamma(\mathbf{r}) = \Gamma_0 + \Gamma_i = \Gamma_0 + \frac{1}{2}\Delta_0[1 - f^2(\mathbf{r})]. \tag{1}$$

We adopt our conductivity model from Refs. [42,43], with one change. Instead of using a single pair-breaking scattering rate for the whole sample, we use a position-dependent rate $\Gamma(\mathbf{r})$ that is set by the local order parameter (see Eq. (1)). From $\overline{\Delta}(\mathbf{r})$ and $\Gamma(\mathbf{r})$, we obtain $\sigma(\mathbf{r}, \omega)$ as described in Sec. S3 of the Supplemental Material [SM]. The local dielectric function follows as $\varepsilon_{Nb}(\mathbf{r}, \omega) = \varepsilon_\infty + i\sigma(\mathbf{r}, \omega)/\varepsilon_0\omega$ ($\varepsilon_\infty = 4$, see SM Sec. S3) and is shown in Fig. 1(c) for several applied fields. These results imply that, as $f \to 0$ on the vortex axis, the local conductivity crosses over continuously to its normal-metal state value. We note that our conductivity model does not capture the dissipation associated with vortex motion, a limitation that is discussed at the end of the paper.

Next, we compute NFRHT using fluctuational electrodynamics and RCWA by employing the approach described in Refs. [13,45] using the open-source package MESH [45] that was co-developed by one of us. In this approach, the heat transfer coefficient $h(d, T)$ is given by [13,46]:

$$h(d, T) = \int_0^\infty \frac{d\omega}{2\pi} h(\omega) = \int_0^\infty \frac{d\omega}{2\pi} \frac{\partial \Theta(\omega, T)}{\partial T} \int_{\mathrm{BZ}} \frac{d^2\mathbf{k}}{(2\pi)^2} \mathcal{T}(\omega, d, \mathbf{k}), \tag{2}$$

where $\Theta(\omega, T) = \hbar\omega/\left[e^{\hbar\omega/k_B T} - 1\right]$ is the mean energy of a Planck oscillator at temperature $T$, $\mathbf{k}$ is the wave vector parallel to the surfaces, and $\mathcal{T}(\omega, d, \mathbf{k})$ is the transmission probability of the electromagnetic modes, summed over polarizations and diffraction orders, while BZ denotes the first Brillouin zone associated with the vortex lattice. Figure 2(a) shows our RCWA (MESH) results (black circles) for the heat-transfer coefficient ($h$) at $d = 100$ nm and $T = 4.2$ K as a function of the applied field ($H$). In the Meissner state, when $H < H_{c1}$ (the lower critical field), $h$ is independent of $H$ and $h_{\text{Meissner}} = 6.2 \times 10^{-3}$ W m$^{-2}$ K$^{-1}$. When $H > H_{c1}$, vortex lines enter Nb and the averaged flux density $\bar{B}$ increases from zero with a vertical tangent, as expected for a second-order transition in the demagnetization-free geometry assumed here (see inset of Fig. 2(a)). As $H$ increases, $h$ rises monotonically by a factor of 2 at $\bar{B} = 30$ mT ($a_0 = 280$ nm), 5 at 95 mT (160 nm), and 13 at 220 mT (104 nm). At $\mu_0 H_{c2} = 0.79$ T, both superconductors become normal, and $h$ reaches its normal-state value $h_N = 1.47$ W m$^{-2}$ K$^{-1}$, implying a switching ratio $h_N/h_{\text{Meissner}} \approx 237$.

The results in Fig. 2(a) raise the question: why does the heat-transfer coefficient $h$ increase monotonically with the applied field $H$? To address this question, we show in Fig. 2(b) the spectral heat flux ($h(\omega)$), corresponding to our RCWA analysis, for a few values of $\bar{B}$ (ranging from the Meissner state to the normal state). It can be seen that in the Meissner state $h(\omega)$ is greatly suppressed for photon energies below the pair-breaking threshold $2\Delta_0 \approx 2.6$ meV, the minimum energy required to break a Cooper pair, and $h(\omega)$ peaks at ~3.3 meV. As $\bar{B}$ increases the sub-gap flux increases rapidly and grows from 16% in the Meissner state to ~41% of the total flux at $\bar{B} =$ 25 mT to ~95% of the total flux at $\bar{B} = 698$ mT. Above $2\Delta_0$ the curves approach the normal-state spectrum as at these photon energies pairs can be broken everywhere in the cell, so the absorption is not affected by the applied field. This analysis shows that the 237-fold rise in Fig. 2(a) is due to a continuous shift of spectral weight into the sub-gap region upon application of the magnetic field.

Next, we explored whether the computationally intensive RCWA results can be captured by an effective medium theory (EMT) similar to those used to interpret terahertz spectroscopy of the vortex state [42]. In this approach, which we call the normal-core EMT, each vortex is modelled as a normal-state core embedded in a superconducting matrix, giving a uniaxial medium $\varepsilon = diag(\varepsilon_{\|}, \varepsilon_{\|}, \varepsilon_{\perp})$ whose in-plane and out-of-plane components follow from the Maxwell–Garnett rule [47]:

$$\varepsilon_{\|} = \varepsilon_{\mathrm{Nb,s}} \frac{\varepsilon_{\mathrm{Nb,n}}(1+f_{\mathrm{n}}) + \varepsilon_{\mathrm{Nb,s}}(1-f_{\mathrm{n}})}{\varepsilon_{\mathrm{Nb,n}}(1-f_{\mathrm{n}}) + \varepsilon_{\mathrm{Nb,s}}(1+f_{\mathrm{n}})}, \quad \varepsilon_{\perp} = \varepsilon_{\mathrm{Nb,s}}(1 - f_{\mathrm{n}}) + \varepsilon_{\mathrm{Nb,n}} f_{\mathrm{n}} \tag{3}$$

with a normal-region fill fraction $f_{\mathrm{n}} = \bar{B}/B_{\mathrm{c2}}$ (this corresponds to cores of radius $\sqrt{2}\xi$, where $\xi$ is the GL coherence length, see SM, Sec. S5 and Ref. [42]) and $\varepsilon_{\mathrm{Nb,n}}$, $\varepsilon_{\mathrm{Nb,s}}$ are the dielectric functions of the cores and matrix. The cores are assigned the normal-state response, i.e. $f \rightarrow 0$ limit of Eq. (1), while the matrix is assigned a single, field-dependent gap set by the value of the order parameter at the edge of the Wigner–Seitz cell (see Supplemental Material [SM]). The computed $h_{\text{normal-core EMT}}$ between two such uniaxial media (following the approach of Ref. [47]) is shown by the red dashed line in Fig. 2(a) and is found to agree with the $h$ computed using RCWA in the Meissner and normal states but overestimates $h$ by up to a factor of 2.5 in the mixed state (see Fig. 2(a), lower panel). This overestimate arises primarily because the normal-core EMT treats a disk of radius $\sqrt{2}\xi$ as fully normal, whereas in the real lattice the order parameter ($f$) recovers to ~0.7 at that radius. Therefore, only a much smaller region near the axis is strongly gapless. We note that assigning a uniformly reduced gap to the matrix contributes a smaller correction of the opposite sign. Both errors vanish as $\bar{B} \rightarrow B_{\mathrm{c2}}$, as $f \rightarrow 0$ throughout the material (see Supplemental Material [SM], Sec. S5 for a quantitative discussion of this point).

To overcome the limitations of the normal-core EMT, we developed a graded-cell EMT in which each Wigner–Seitz cell is modelled as a uniaxial medium whose dielectric tensor is built directly from the computed $\varepsilon_{\mathrm{Nb}}(r,\omega)$. Specifically, the out-of-plane component is obtained from the area average, $\varepsilon_{\perp}(\omega) = \frac{2}{R^2}\int_0^R \varepsilon_{\mathrm{Nb}}(r,\omega)\, r\, dr$, where $R = 0.525\, a_0$ is the radius of the circular cell of the same area as the Wigner–Seitz cell. The in-plane component $\varepsilon_{\|}(\omega)$, on the other hand, is obtained by discretizing the cell into thin concentric annuli and applying the Maxwell–Garnett relation outward from the core. In this approach each homogenized disk of radius $r_{\mathrm{k}}$ acts as the inclusion for the next annulus $\varepsilon_{\mathrm{Nb}}(r_{\mathrm{k+1}}, \omega)$ at area fraction $(r_{\mathrm{k}}/r_{\mathrm{k+1}})^2$ (see Supplemental Material [SM], Sec. S5). The solid blue line in Fig. 2(a) shows the graded-cell EMT result, which agrees closely with RCWA (MESH), to within 0.5% at every $H$ field above ~105 mT and to within 3% at the most dilute lattice ($a_0 = 482$ nm, $\bar{B} \approx 10.3$ mT). The close agreement between RCWA and graded-cell EMT models is because the vortex spacing (55–482 nm) is below the lateral wavelength ~$2\pi d \approx 0.6$ µm of the near-field modes that carry the flux (i.e. $k_{\parallel} \lesssim 1/d$), which homogenizes the lattice. The graded-cell EMT approach gives better results when compared with the normal-core EMT approach as the graded profile removes the artificial normal/superconducting boundary assumed in the normal-core model. We note that the $H$-field-driven enhancement in our system is dissipative and broadband rather than resonant. Past work on NFRHT in metasurfaces has reported resonant enhancements associated with surface modes, which appear as a bright dispersive branch in the transmission $\mathcal{T}(\omega, k_{\parallel})$ [13,47]. We find that no such branch arises in the transmission for our vortex metasurface (see Supplemental Material [SM], Sec. S6). Instead, NFRHT in the vortex metasurface is carried by a broad range of evanescent wavevectors with $k_{\parallel}$ between $\frac{0.2}{d}$ and $\frac{1}{d}$ in all states (Meissner, mixed and normal) as the applied $H$ field increases the sub-gap NFRHT without introducing a resonance.

Subsequently, we explored how NFRHT depends on the gap size $d$ and temperature $T$. Figure 3(a) presents $h$ as a function of $d$ (10 nm to 5 µm) computed with the graded-cell EMT (the Meissner- and normal-state curves, and hence the contrast, involve no homogenization whereas the mixed-state curves below $d \sim a_0/2\pi$, and all curves below $d \sim 30$ nm, where $k_{\parallel}\,\ell \approx \ell/d$ is no longer small, should be regarded as indicative) for the normal state, at three mean flux densities ($\bar{B}$ = 133, 493 and 698 mT), and in the Meissner state. In all cases $h$ decreases rapidly with increasing gap, but at different rates, so that the switching contrast ($h_{\mathrm{N}}/h_{\mathrm{Meissner}}$, Fig. 3(b)) changes from ~237 at 100 nm to ~884 at 1 μm to a broad maximum of ~1500 at $d \approx 3.5$ μm, beyond which it decreases. This behavior arises from two distinct effects. Below a few μm, NFRHT is dominated in both the Meissner and normal states by $s$-polarized evanescent waves with $k_{\parallel} \lesssim 1/d$ [2,40]. Reducing the gap size adds evanescent channels of ever larger in-plane wavevectors ($k_{\parallel}$), but there exists a cutoff value for wavevectors beyond which the surface no longer absorbs [40]. For a normal metal this cutoff is determined by the inverse of the skin depth, $\delta = (2\rho_0/\mu_0\omega)^{1/2}$, which ranges from ~500 nm to ~130 nm across the thermal window at 4.2 K. For the superconductor it is the inverse of the magnetic penetration depth, $\lambda \approx 88$ nm, which is nearly frequency-independent for ħω well below $2\Delta_0$. The normal-state flux therefore saturates once $d$ falls below ~$\delta$, while the Meissner-state flux keeps growing down to gaps of order $\lambda$ (inset of Fig. 3(b)). This explains why the switching contrast falls at small gaps.

To understand the behavior at larger gaps, a comparison of the relative contributions of the s- and p-polarized channels is needed (SM, Sec. S7). With increasing gap size the s-polarized flux decays in both states, and the Meissner-state flux decays faster over the range studied (SM, Sec. S7). From the s channel alone, the contrast would continue to rise well beyond the gap at which it is observed to peak. What limits the rise and produces the observed peak is the p-polarized flux. This channel is carried by waves close to the light line, which are attenuated only weakly across the gap, and we find that its flux falls only as $d^{-1}$ over the whole range (SM, Sec. S7). It contributes, however, only at frequencies where the surface absorbs, because a lossless surface reflects these waves completely. In the normal state the p channel stays negligible relative to the s channel over the range studied. In the Meissner state, by contrast, the s channel is weak because sub-gap absorption is small (see Im $\varepsilon_{Nb}$ in Fig. 1(c)), whereas the p channel is carried by photons above $2\Delta_0$, which the superconductor absorbs comparably to a normal metal. The p channel therefore overtakes the s channel at $d \approx 6$ µm. As the p channel becomes a significant fraction of the Meissner-state flux, that flux stops falling steeply, and the contrast passes through a maximum of about 1500 at $d \approx 3.5$ µm (SM, Sec. S7).

Finally, we explore the temperature dependence of the switching ratio at a 100 nm gap size by computing the switching ratio from 2 K to 8.5 K ($T_c$ for our sample) with the graded-cell EMT. The computed results are shown in Fig. 3(c), where the switching contrast increases monotonically from 237 at 4.2 K to $5.1 \times 10^3$ at 3 K and $6.7 \times 10^5$ at 2 K, roughly one to two decades per kelvin of cooling, while it decreases monotonically upon heating to $T_c$. This behavior can be understood by noting that the Meissner-state flux is exponentially activated across the gap, i.e., $h_{\mathrm{Meissner}} \propto \exp(-2\Delta_0/k_B T)$ for $k_B T \ll \Delta_0$, whereas the normal-state flux varies only weakly with $T$. The inset of Fig. 3(c) explains the origin of the large contrast at 2 K. As can be seen from the inset, at 2 K the normal-state spectrum features a broad peak at 0.36 meV reaching 0.63 W m$^{-2}$ K$^{-1}$ meV$^{-1}$, while the Meissner state has almost no heat transfer at those energies. Instead, the spectral heat flux in the Meissner state is a plateau at ~$10^{-7}$ W m$^{-2}$ K$^{-1}$ meV$^{-1}$, almost seven orders of magnitude below the normal-state peak, and two thirds of its (much smaller) total heat flux comes from a narrow peak at 3.1 meV, just above $2\Delta_0$. This behavior can be understood by noting that cooling to 2 K lowers the thermal photon energies (the normal-state peak moves from 0.69 meV at 4.2 K to 0.36 meV at 2 K) while the gap stays nearly fixed, so the overlap between the thermal spectrum and the superconductor's absorption shrinks exponentially, thus increasing the switching contrast.

In this paper we assumed that the vortex cores in the mixed state are fixed. In practice the vortices may move in response to electromagnetic fluctuations. A time-dependent treatment including flux flow [33,42,48] could be employed to refine the expected response within the mixed state. We note that the switching ratio predicted in this work is independent of the model assumed for the vortex phase because it is set by the Meissner and normal states, whose heat transfer involves no vortices and is therefore unaffected by vortex motion. We also note that pure Nb is in the clean limit, with $\ell \sim 1$ µm, so that its electrodynamics is nonlocal [40], and is type-II/1 [49], therefore it offers a distinct regime for exploring NFRHT. Finally, in samples with pinning strong enough to trap flux at zero applied field [50], the enhanced heat transfer of the mixed state would persist after the field is removed. Such remanent vortex lattices would allow latching thermal switches that hold their state without a sustained field and could be reset by an AC demagnetization sequence, offering opportunities for creating novel cryogenic thermal tuning devices.

To summarize, our work shows that vortex matter provides a self-assembled, magnetically tunable metasurface that, unlike lithographically patterned metasurfaces, requires no micro- or nanofabrication and allows continuous tuning of NFRHT at fixed temperature by more than two orders of magnitude at 4.2 K, and by more than five at 2 K. These predictions can be tested with the experimental tools currently available [22,25] for probing NFRHT across nanoscale gaps at cryogenic temperatures.

**Acknowledgements:** S.S. acknowledges useful discussions with Dr. Yuxuan Luan and Prof. Edgar Meyhofer of the Univ. of Michigan. We acknowledge the use of the Great Lakes HPC cluster at the Univ. of Michigan. S.F. acknowledges funding support from the Department of Energy (Grant No. DE-FG02-07ER46426).

**Author contributions:** S.F. and P.R. conceived the research. S.S. developed the graded-cell EMT model and the order-parameter-based dielectric function under the guidance of S.F. S.S. performed all calculations and analyzed the data under the guidance of S.F. and P.R. S.S. prepared the first draft of the manuscript, which was edited and finalized by P.R. and S.F. All authors discussed the results and approved the manuscript.

**Data Availability Statement:** Data and code generated during the study are available from the authors on reasonable request.

†Corresponding author
shanhui@stanford.edu

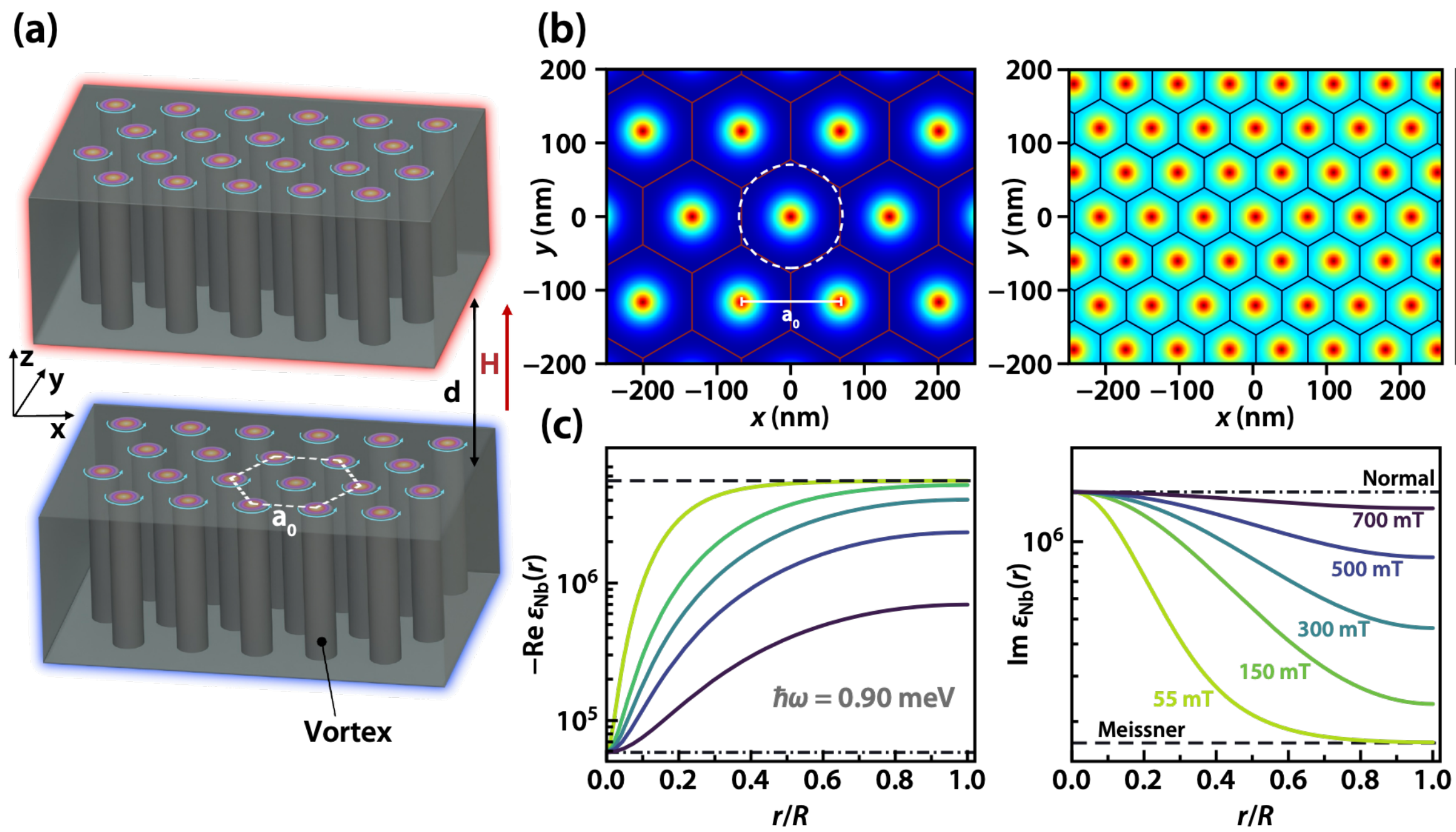

**FIG. 1.** (a) Schematic of the vortex metasurface formed by two Nb bodies in the mixed state (not to scale). Their surfaces are parallel and separated by a vacuum gap *d*, and the field *H* is applied along *z*, perpendicular to the surfaces. The bodies are long along *z*, so that demagnetization can be neglected. Each body carries a triangular lattice of vortices of spacing $a_0$ embedded in the superconducting matrix. Each vortex features a circulating current that is indicated by a cyan arrow. The lattices are assumed to be aligned with respect to each other. The upper (red) and lower (blue) bodies are held at different temperatures. (b) The computed Ginzburg–Landau order parameter $f(r)$ of the vortex lattice at $\mu_0 H$ = 150 mT (left, $a_0$ = 134 nm) and 500 mT (right, $a_0$ = 70 nm), on the same scale. Here $f$ = 0 on the vortex axes and between them $f$ recovers to ≈1 at 150 mT and ≈0.8 at 500 mT. The solid lines are the hexagonal Wigner–Seitz cells. The dashed circle in the left panel, of radius $R$ = 70 nm, has the same area as the Wigner–Seitz cell and is the circular cell used in the calculations of (c) and of Fig. 2. (c) Local dielectric function of Nb at $\hbar\omega$ = 0.90 meV, with -Re($\varepsilon_{Nb}$) (left panel) and Im($\varepsilon_{Nb}$) (right panel), against the distance $r$ from the vortex axis normalized to the circular-cell radius $R$, for five applied fields ($R$ = 158, 70, 48, 37 and 31 nm for applied $\mu_0 H$ = 55, 150, 300, 500 and 700 mT, respectively). The dashed and dash-dotted lines are the Meissner and normal states, respectively.

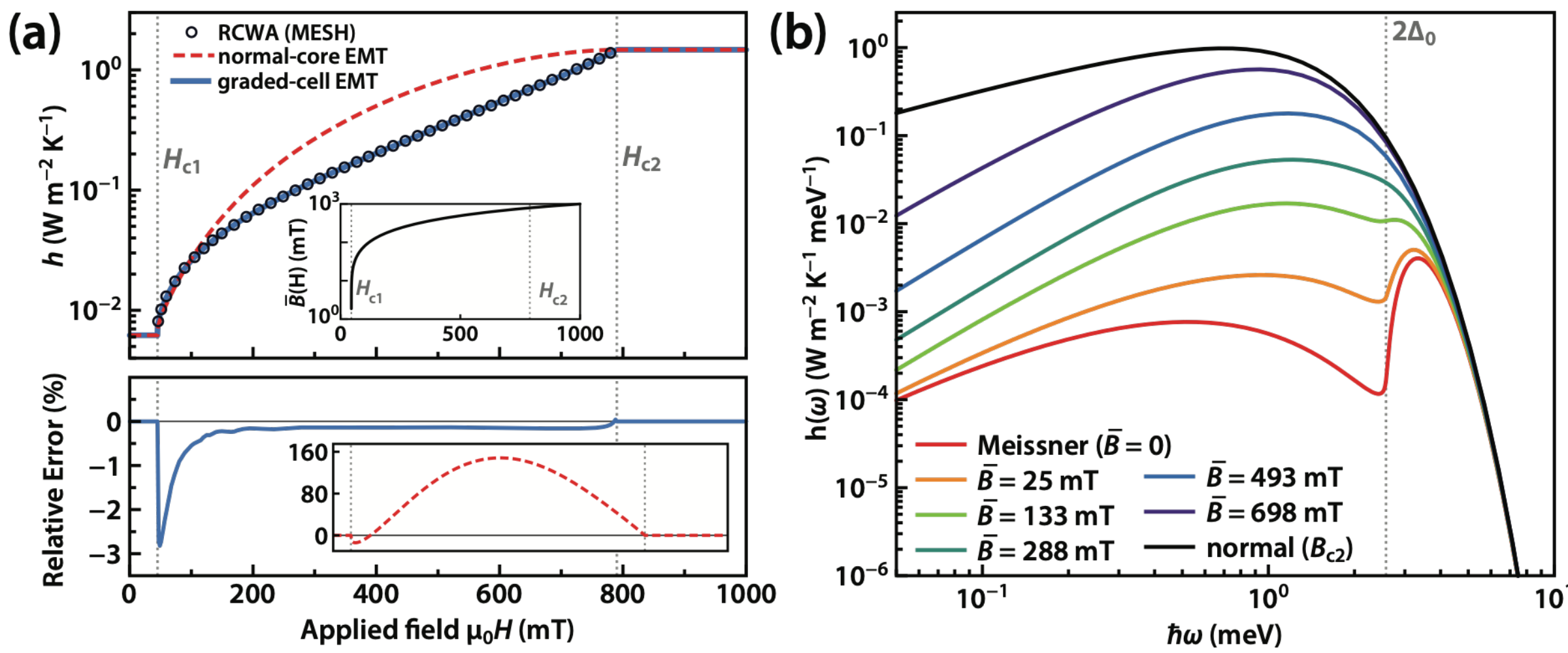

**FIG. 2.** (a) Top panel shows the near-field heat-transfer coefficient ($h$) of the vortex metasurface at $d$ = 100 nm and $T$ = 4.2 K as a function of the applied $H$ field. Circles: RCWA (MESH) calculation for the exact lattice. Solid line: the graded-cell effective medium theory (graded-cell EMT). Dashed line: the two-phase normal-core Maxwell–Garnett model. Inset of the top panel shows the computed $\bar{B}$ vs. $\mu_0 H$. The bottom panel shows the relative error 100× ($h_{\text{graded-cell EMT}}$ - $h_{\text{RCWA}}$)/$h_{\text{RCWA}}$. Inset shows the relative error for the normal-core EMT. The graded-cell EMT reproduces the exact lattice to within 0.5% for applied $H \gtrsim$ 105 mT and 3% down to the most dilute lattice. (b) Spectral heat flux, $h(\omega)$, for different values of $\bar{B}$ (25, 133, 288, 493 and 698 mT), with the Meissner (red) and normal-state (black) references computed using RCWA. The dotted line marks the pair-breaking threshold $2\Delta_0$ = 2.6 meV.

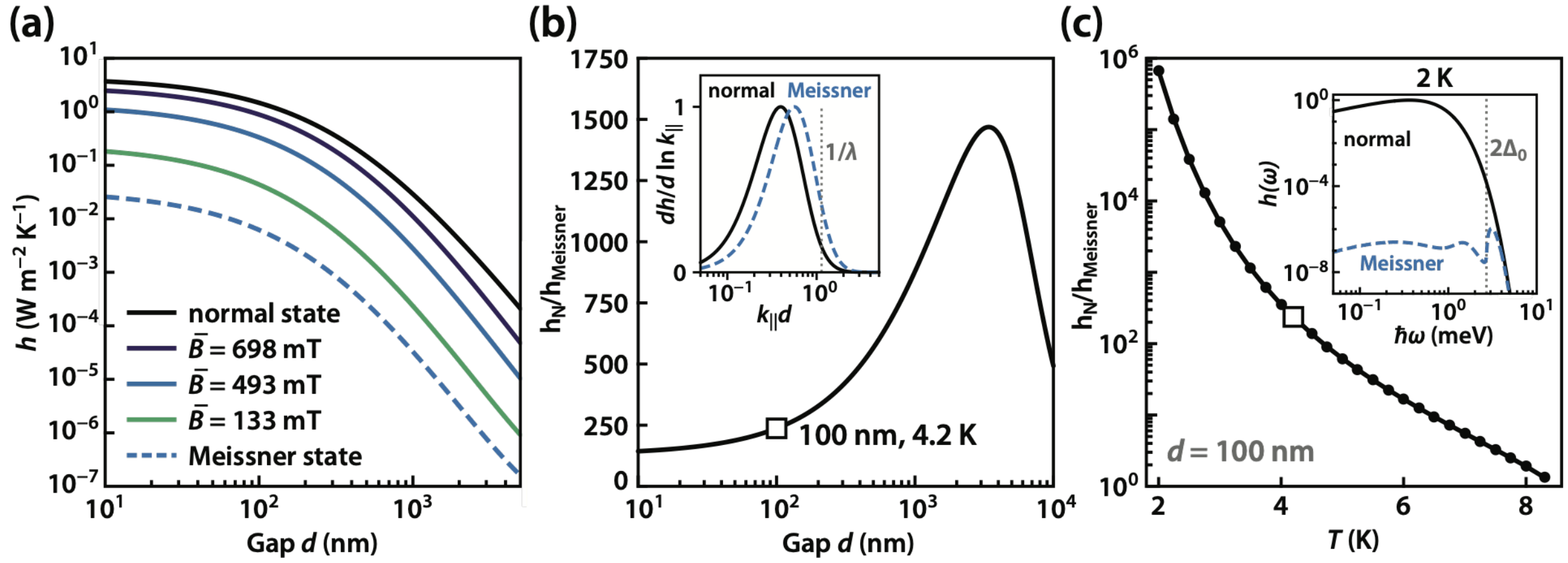

FIG. 3. (a) Gap dependence of the heat-transfer coefficient at $T = 4.2$ K, computed using graded-cell EMT, in the normal state, at three magnetic inductions ($\bar{B}$ = 133 mT, 493 mT, 698 mT), and in the Meissner state. (b) Switching contrast $h_N/h_{Meissner}$ as a function of the gap size. The square (in panels b and c) marks the (100 nm, 4.2 K) operating point considered in Fig. 2. Inset shows the wavevector distribution $dh/d\ln k_\parallel$ of the heat flux in the normal (solid) and Meissner (dashed) states at $d = 100$ nm. (c) Switching contrast as a function of temperature at $d = 100$ nm. Inset shows the spectral heat flux of the normal (solid) and Meissner (dashed) states at 2 K. The dotted line marks the energy corresponding to $2\Delta_0$.

**Supplemental Material**

# Vortex matter as a magnetically programmable metasurface for tuning near-field radiative heat transfer

Siddharth Sangi[1], Pramod Reddy[2], and Shanhui Fan[3,†]

[1]*Huron High School, Ann Arbor, MI 48105, USA.*

[2]*Department of Mechanical Engineering, University of Michigan, Ann Arbor, MI 48109, USA.*

[3]*Department of Electrical Engineering, Ginzton Laboratory, Stanford University, Stanford, CA 94305, USA.*

**Table of contents**

## S1. Material parameters of oxygen-doped niobium

All the material parameters for the Nb sample considered in this work were obtained from measurements reported in Ref. [1]. Specifically, Table I of Ref. [1] gives, for several oxygen-doped Nb samples, the measured resistivity $\rho_n$, the resistively determined critical temperature $T_c$, the thermodynamic critical field $H_c$ (obtained from the area under the ascending magnetization curve), and the upper critical field ($H_{c2}$) at which the magnetization ($M$) vanishes. Ref. [1] also lists the first-penetration field $H_{c1}^{\exp}$ (labelled $H_{FP}$ in Ref. [1]), which is the field at which flux first enters the sample and, because of the surface barrier, lies above the thermodynamic lower critical field ($H_{c1}$). The data in Ref. [1] (a subset of which is listed below in Table S1) show that $\rho_n$ rises by ~5.2 μΩ cm per at.% O. For $\rho_0 = 5\,\mu\Omega\,\mathrm{cm}$ (the value assumed for our sample) the corresponding oxygen concentration is 0.96 at.%. The bottom row of Table S1 gives $T_c$, $H_c$, $B_{c2}$ and $H_{c1}^{\exp}$ at this composition, obtained by linear interpolation in oxygen content between the 0.70 and 1.52 at.% O samples of Ref. [1] and rounded to two significant figures. The resistivity in that row is the assumed $\rho_0$, not an interpolated value.

*TABLE S1. Data from Ref. [1] at 4.2 K (SI units) and the interpolation to $\rho_0 = 5\,\mu\Omega\,cm$.*

| Sample | $\rho_n$ ($\mu\Omega$cm) | $T_c$ (K) | $\mu_0 H_c$ (mT) | $B_{c2}$ (T) | $\mu_0 H_{c1}^{\exp}$ (mT) |
|---|---|---|---|---|---|
| Nb + 0.70 at.% O | 3.9 | 8.78 | 136.0 | 0.700 | 58 |
| Nb + 1.52 at.% O | 8.2 | 8.04 | 112.5 | 0.967 | 35 |
| interpolated, 0.96 at.% O | 5.0 | 8.5 | 130 | 0.79 | 51 |

In the Letter, we assume that Nb contains ~0.96 at.% interstitial oxygen, resulting in a residual resistivity (impurity-limited resistivity at low temperatures) $\rho_0$ = 5 μΩ cm (Table S1). This Nb has a mean free path $\ell \approx 7.5$ nm, critical temperature $T_c$ = 8.5 K, Ginzburg–Landau parameter $\kappa \approx 4.3$, lower critical field $\mu_0 H_{c1}$ = 45 mT and upper critical field $\mu_0 H_{c2} \equiv B_{c2}$ = 0.79 T, all at 4.2 K. Of these, $T_c$ and $B_{c2}$ are read directly from Table S1. Below we explain how $\kappa$, $\mu_0 H_{c1}$, $\xi$, $\lambda$ and the gap $\Delta(T)$ are obtained from the tabulated fields, and how $\ell$ is obtained from $\rho_0$.

***Estimation of the Ginzburg–Landau parameter (κ)***

Below $T_c$ the thermodynamic field $H_c$, the lower critical field $H_{c1}$ and the upper critical field $H_{c2}$ are no longer related through a single Ginzburg–Landau parameter $\kappa$. Past work [2,3] has shown that the values inferred from $H_{c2}$, from the magnetization slope at $H_{c2}$ and from $H_{c1}$ (i.e., $\kappa_1$, $\kappa_2$ and $\kappa_3$) coincide only at $T_c$ and differ for $T < T_c$. In this work, we define the parameter $\kappa$ used in our implementation of Brandt's approach [4] as:

$$\kappa \equiv \kappa_1 = \frac{B_{c2}}{\sqrt{2}\mu_0 H_c} = \frac{790\ \mathrm{mT}}{\sqrt{2} \times 130\ \mathrm{mT}} = 4.3. \quad \text{(S1)}$$

Next, we obtain $H_{c1}$ from the Ginzburg–Landau-based expression of Ref. [4]:

$$\frac{H_{c1}}{H_{c2}} = h_{c1}(\kappa) = \frac{\ln\kappa + \alpha(\kappa)}{2\kappa^2}, \alpha(\kappa) = 0.49693 + \exp[-0.41477 - 0.775\ln\kappa - 0.1303\ln^2\kappa], \quad \text{(S2)}$$

which for the above value of $\kappa$ gives $\mu_0 H_{c1} = 0.0573 \times 790\ \mathrm{mT} \approx 45\ \mathrm{mT}$. We note that the first-penetration field of 51 mT (from Table S1) lies 13% above this value since flux entry is delayed by the Bean–Livingston surface barrier, so $H_{c1}^{\mathrm{exp}}$ is an upper bound on the thermodynamic $H_{c1}$. Therefore, a value of $\mu_0 H_{c1} = 45\ \mathrm{mT}$ is used throughout this work.

***Estimation of the coherence length (ξ) and penetration depth (λ)***

The Ginzburg–Landau identities $\mu_0 H_{c2} = B_{c2} = \Phi_0/(2\pi\xi^2)$ and $\lambda = \kappa\xi$ [5] give:

$$\xi = \sqrt{\frac{\Phi_0}{2\pi B_{c2}}} = 20.4\ \mathrm{nm}, \qquad \lambda = \kappa\xi = 88\ \mathrm{nm}. \quad \text{(S3)}$$

***Estimation of the mean free path ($\ell$) and a description of the dirty-limit criteria***

The electron mean free path follows from the Drude relation $\rho\ell = mv_F/ne^2$, which for Nb is $\rho_0\ell = 3.75 \times 10^{-12}\ \Omega\,\text{cm}^2$ [6]:

$$\ell = \frac{\rho_0\ell}{\rho_0} = 7.5\ \text{nm}. \tag{S4}$$

Hence $\ell/\xi_0 = 0.19$, with $\xi_0 = 39$ nm the clean-limit coherence length [6]. Therefore, the Nb considered in this work is in the dirty limit, $\ell \ll \xi_0$, in which impurity scattering rather than $\xi_0$ sets the superconducting length scales, $\xi \simeq 0.855\sqrt{\xi_0\ell/(1-t)}$ ~20.6 nm (in good agreement with the value obtained in Eq. (S3)), where $t = T/T_c$ [5]. Independently, $\ell/\lambda = 0.09$ and $\ell/d = 0.08$ (with $d$ = 100 nm the vacuum gap, the relevant wavevector being $q \approx 1/d$), so $\ell$ is also short compared with the scales over which the fields vary. Given these conditions the electrodynamics describing our Nb is expected to be local [7], justifying our approach for modelling NFRHT.

***Estimation of the superconducting gap $\Delta(T)$***

The conductivity model of Sec. S3 requires the zero-field gap $\Delta_0 = \Delta(T)$, at every $T$ between 2 K and $T_c$. Since this gap is not reported in Ref. [1], we obtain it from $T_c$ in two steps. First, the zero-temperature gap of Nb follows from $2\Delta(T=0)/k_BT_c = 3.67$ [6]. We assume that this ratio is unchanged by the oxygen impurities, which are nonmagnetic and therefore not pair-breaking [8]. Second, the temperature dependence is given by [9]:

$$\Delta(T) = 1.835\,k_BT_c\sqrt{1-t}\,(0.9663 + 0.7733\,t), \tag{S5}$$

which, with $t = \frac{T}{T_c} =$ 0.494 and $1.835\,k_BT_c = 1.344$ meV ($T_c = 8.5$ K), gives $\Delta_0 \equiv \Delta(4.2\ \text{K}) =$ 1.289 meV. This is the value that we use in Sec. S3 and quoted (rounded to 1.3 meV) in the Letter.

**S2. Approach to evaluating the mean flux density, lattice spacing and order parameter of the vortex lattice**

The approach described in the main text requires the mean flux density $\bar{B}(H)$, the vortex spacing $a_0$ and the order parameter $f(r) = |\psi(r)|/\psi_\infty$ (see Fig. 1(b)) for various applied $H$ fields. All these three quantities are obtained from the Ginzburg–Landau vortex-lattice solution of Ref. [4] in the circular-cell approximation (Ref. [4], Appendix A), which requires as inputs $\kappa = 4.3$, $B_{c2} = 0.790\ T$ and $\lambda = 88$ nm (all from Sec. S1). The order-parameter profile shown in Fig. 1 is tiled over the triangular lattice, and the value at the cell boundary is read from $f(r)$ at the equal-area radius $R$. The circular cell reproduces the free energy of the exact triangular lattice to better than 0.2% [4], so the use of the circular cell introduces negligible error. The vortex spacing and the radius $R$ of the equal-area circular cell are given by:

$$a_0 = \sqrt{\frac{2\Phi_0}{\sqrt{3}\bar{B}}}, \quad R = \sqrt{\frac{\Phi_0}{\pi\bar{B}}} = 0.525\, a_0 \tag{S6}$$

We note that Ginzburg–Landau theory is strictly valid only near $T_c$. At 4.2 K (reduced temperature $t = T/T_c = 0.494$) we use it as an approximation to obtain $f(r)$, $\bar{B}$ and $a_0$. The reliability of this approximation is confirmed by the fact that our approach reproduces the measured magnetization to within ~10% (for the 0.70 at.% O sample in [1]) across the mixed state with no adjustable parameters. The computed values of $\bar{B}$ and $a_0$ (for our sample) over the field range of interest are given in Table S2.

*TABLE S2. Computed mean flux density and vortex spacing versus applied H field for Nb + 0.96 at.% O at 4.2 K ($\kappa = 4.3$, $B_{c2} = 0.790$ T, $\mu_0 H_{c1} = 45$ mT).*

| **$\mu_0$H (mT)** | 50 | 60 | 80 | 100 | 150 | 200 | 300 | 500 | 700 |
|---|---|---|---|---|---|---|---|---|---|
| **$\bar{B}$ (mT)** | 18.1 | 33.0 | 57.2 | 79.5 | 132.7 | 184.8 | 287.8 | 493.0 | 697.8 |
| **$a_0$ (nm)** | 363 | 269 | 204 | 173 | 134 | 114 | 91 | 70 | 58 |

## S3. Approach to evaluate the local optical conductivity of the vortex lattice

In the main text we stated that the local dielectric function of Nb, $\varepsilon_{\mathrm{Nb}}(r,\omega)$, is built from the position-resolved conductivity $\sigma(r,\omega)$, evaluated at each point of the vortex lattice by adapting the approach of Refs. [10,11]. Below, we briefly describe the approach of Refs. [10,11] and then our adaptation of it. For consistency with Refs. [10,11] all energies are in meV and $\hbar = 1$.

***Summary of the approach of Refs. [10,11]***

Refs. [10,11] express the optical conductivity of a homogeneous superconductor in terms of a pair potential $\bar{\Delta}$, a pair-breaking rate $\Gamma$, and a pair-conserving rate $\Gamma_{\mathrm{s}}$. In Ref. [11] the magnetic field dependence enters through $\Gamma$ alone, which is fitted to each measured spectrum; the pair-correlation gap $\bar{\Delta}$ then follows from the relation $\bar{\Delta} = \sqrt{[\Delta_0(\Delta_0 - 2\Gamma)]}$, where $\Delta_0$ is the gap in the absence of pair breaking, and $\Gamma_{\mathrm{s}}$ is independent of the field. Refs. [10,11] employ renormalized functions:

$$\widetilde{\Delta}(E) = \overline{\Delta}\left(1 + \frac{i\Gamma}{E}\right), \quad Z(E) = \left(1 + \frac{i\Gamma}{E}\right)\left(1 + \frac{i\Gamma_s}{\Omega(E)}\right), \quad \Omega(E) = \sqrt{(E + i\Gamma)^2 - \overline{\Delta}^2}, \qquad \text{(S7)}$$

where $\widetilde{\Delta}(E)$ is the energy-dependent renormalized gap and $\bar{\Delta}$ the pair-correlation gap introduced above. Following S. B. Nam [12] three additional functions:

$$n = (\omega + i\Gamma)/\Omega(\omega),\ p = \overline{\Delta}/\Omega(\omega) \text{ and } \epsilon = \Omega(\omega) + i\Gamma_s, \qquad \text{(S8)}$$

together with two auxiliary functions

$$H_1(x,y) = \frac{1 + n(x)n^*(y) + p(x)p^*(y)}{2[\epsilon^*(y) - \epsilon(x)]}, \qquad H_2(x,y) = \frac{1 - n(x)n(y) - p(x)p(y)}{2[\epsilon(y) + \epsilon(x)]}, \qquad \text{(S9)}$$

are introduced from which the optical conductivity $\sigma(\omega)$, at photon energy $\omega$ is obtained using (note that in Eqs. S8 and S9 the argument $\omega$ denotes the quasiparticle energy, identical to $E$ in Eq. (S7) since $\hbar = 1$):

$$\sigma(\omega) = \frac{iD_0}{\omega}\int_{-\infty}^{+\infty} dx\{[f_e(x+\omega) - f_e(x)]H_1(x+\omega, x) - f_e(x)[H_2(x+\omega, x) + H_2^*(x, x-\omega)]\}, \quad (S10)$$

Here, $f_e$ is the Fermi function (not to be confused with the order parameter $f$(r)) and $D_0 = ne^2/m$ corresponds to the normal state and is related to the dc conductivity by $\sigma_n = D_0/2\Gamma_s$. It is clear from the above equations that the necessary inputs for evaluating the conductivity are $\Delta_0$, $\Gamma$, $\Gamma_s$ and $\sigma_n$.

***Adaptation of the approach of Ref.* [11] *to evaluate the local electrical conductivity* $\sigma(r,\omega)$**

While Ref. [11] treats the superconductor as homogeneous and obtains the field dependence from a scattering rate Γ fitted to each spectrum, we seek to resolve the conductivity locally. We achieve this through the Ginzburg–Landau order parameter $f(r)$ of Sec. S2, which enables the evaluation of the local pair correlation gap via: $\bar{\Delta}(r) = f(r)\Delta_0$. Further, we relate $\Gamma(\mathbf{r})$ to $f(\mathbf{r})$ via (see main manuscript Eq. 1):

$$\Gamma(\mathbf{r}) = \Gamma_0 + \frac{1}{2}\Delta_0[1 - f^2(\mathbf{r})], \quad (S11)$$

where $\Gamma_0$ is the intrinsic (zero-field) pair-breaking scattering rate, discussed in more detail below. In our approach the zero-field gap is $\Delta_0 = 1.3$ meV at 4.2 K (Sec. S1); the intrinsic scattering rate is $\Gamma_0 = 0.01$ meV (discussed below); the normal-state conductivity is $\sigma_n = 1/\rho_0 = 2.0 \times 10^7$ S/m, from the residual resistivity, and the pair-conserving scattering rate follows from the Drude relation as $\Gamma_s = \hbar\varepsilon_0\omega_p{}^2/2\sigma_n = 11.3$ meV with $\omega_p = 8.8 \times 10^{15}$ s$^{-1}$ [13,14].

The local conductivity ($\sigma(r,\omega)$) is then obtained point by point. At each position $r$, the homogeneous-superconductor expressions (S7)–(S10) are evaluated with the local gap $\bar{\Delta}$(r) and local scattering rate Γ(r) in place of the uniform $\bar{\Delta}$ and Γ of Ref. [11], while $\Gamma_s$ and $\sigma_n$ are the same everywhere. Finally, the local dielectric function is given by $\varepsilon_{Nb}(r,\omega) = \varepsilon_\infty + i\sigma(r,\omega)/\varepsilon_0\omega$, with $\varepsilon_\infty = 4$ (see Ref. [15]).

***Intrinsic pair-breaking rate***

We note that $\Gamma_0$ is the zero-field pair-breaking scattering rate (i.e., the pair-breaking rate in the absence of an applied *H* field, intrinsic to the material), the only quantity in the model not fixed by Ref. [1] or by Ginzburg–Landau identities. In the manuscript, we use $\Gamma_0 = 0.01$ meV, which is at the lower end of values reported for Nb [16]. Since the contrast falls monotonically with $\Gamma_0$ (Table S3), here we quantify the impact of increasing $\Gamma_0$ on the switching contrast. We find that at 4.2 K, raising $\Gamma_0$ from 0.005 to 0.05 meV changes the contrast at *d* = 100 nm from 243 to 191, about 20% over a factor of ten in $\Gamma_0$. Variation of $\Gamma_0$ has a larger impact at low temperatures, and results in an ~8.3 fold reduction in the switching contrast, although the switching contrast still continues to be of the order of $10^5$.

*TABLE S3. Sensitivity of the switching contrast $h_N/h_{Meissner}$ at d = 100 nm to $\Gamma_0$.*

| **$\Gamma_0$ (meV)** | **0.005** | **0.01** | **0.02** | **0.03** | **0.05** |
|---|---|---|---|---|---|
| **T = 4.2 K** | 243 | 237 | 225 | 213 | 191 |
| **T = 3 K** | $5.4\times10^3$ | $5.1\times10^3$ | $4.5\times10^3$ | $4.0\times10^3$ | $3.0\times10^3$ |
| **T = 2 K** | $8.3\times10^5$ | $6.7\times10^5$ | $4.0\times10^5$ | $2.4\times10^5$ | $1.0\times10^5$ |

**S4. Rigorous coupled wave analysis (RCWA+MESH) calculations and convergence**

The near-field heat-transfer coefficient ($h$) between the two Nb bodies was computed with the software MESH, which was co-developed by one of us [17]. MESH offers a scattering-matrix implementation of fluctuational electrodynamics for periodic, layered media, which evaluates Eq. (2) of the main text by rigorous coupled-wave analysis (RCWA). We summarize here how the vortex state was represented in MESH and the numerical settings that we used. The method and software are described in detail in Ref. [17].

We represent each superconductor as a semi-infinite medium whose in-plane permittivity $\varepsilon_{\mathrm{Nb}}(r, \omega)$ is periodic on the triangular vortex lattice in the $x$, $y$ -directions (see Fig. 1 of the manuscript) and is independent of the coordinate $z$ along the vortices (the surface normal). We represented the triangular lattice of vortices as a rectangular supercell of dimensions $a_0 \times \sqrt{3}a_0$ containing two vortices. Each vortex is built from a set of concentric rings whose permittivities are the values of $\varepsilon_{\mathrm{Nb}}(r, \omega)$ at the local order parameter $f(r)$ that we computed using the Ginzburg–Landau approach (see Sec. S2). The two Nb bodies are assumed to be identical and the vortices are assumed to be in registry across the vacuum gap ($d$ = 100 nm) as this is the thermodynamically stable configuration, since the flux lines thread the gap. MESH constructs the scattering (reflection) operators of the two half-spaces in the plane-wave basis of diffraction orders, propagates them across the gap, and outputs the spectral transmission summed over the diffraction orders and polarizations. Finally, $h$ is calculated from the Brillouin-zone and frequency integrals of Eq. (2).

In our calculations performed using MESH two parameters control the accuracy: 1) the number of reciprocal-lattice vectors (diffraction orders) $N_{\mathrm{G}}$ resolves the in-plane structure of the cell, and 2) the Brillouin-zone integral is evaluated on a uniform *k*-grid in reciprocal space. We systematically scaled both of them with the lattice spacing and varied them as needed until the computed value of $h$ converged. We found that for dilute lattices, whose cell size is large and whose near field couples to many orders, require the largest $N_G$ (up to 254), while the dense lattices, whose Brillouin zone is large, require the finest *k*-grid. The settings that we used are listed in Table S4 below. The data in the table show that doubling either parameter changes $h$ by less than the tolerances quoted there. The frequency integral used 48 points spaced geometrically from 0.05 to 10 meV, sufficient for $h$ to better than 0.2%.

*TABLE S4. Numerical settings for the MESH calculation and the convergence of h under doubling of each parameter (k and $N_G$).*

| lattice | $a_0$ (nm) | $N_G$ | *k*-grid | convergence of $h$ |
|---|---|---|---|---|
| dense (0.9 $B_{c2}$) | 58 | 70 | 36 × 22 | < 0.3% |
| intermediate | 130 | 118 | 16 × 10 | < 0.3% |
| dilute (near $H_{c1}$) | 300 | 254 | 8 × 6 | < 0.5% |

We note that our calculations were validated against the two homogeneous limits (Meissner and normal state), for which MESH reproduces the results obtained using Polder-Van Hove's approach [18]. Specifically, we found that the Meissner and normal states reproduce the values computed from Polder-Van Hove's theory to within 0.1%.

**S5. Description of the graded-cell EMT and comparison to RCWA**

In the graded-cell EMT approach we replace each vortex cell by a homogeneous, uniaxial effective medium whose dielectric tensor is built from the computed $\varepsilon_{\mathrm{Nb}}(r,\omega)$ (see Sec. S3). Here, we describe how the in-plane and out-of-plane components of the dielectric tensor are computed. Since $\varepsilon_{\mathrm{Nb}}(r,\omega)$ varies only in the plane and is invariant along the vortex axis $z$, the optic axis lies along $z$. The out-of-plane component, for which all points of the cell respond in parallel, is given by the area average of the dielectric function ($\varepsilon_{\mathrm{Nb}}(r,\omega)$):

$$\varepsilon_{\perp}(\omega) = \frac{2}{R^2}\int_0^R \varepsilon_{\mathrm{Nb}}(r,\omega)\, r\, dr \tag{S12}$$

where $R = 0.525\, a_0$ is the radius of the circular cell of the same area as the Wigner–Seitz cell. The in-plane component $\varepsilon_{\parallel}$ was obtained by using a coated-cylinder recursive approach. Starting from the vortex axis, the circular cell is divided into $N$ concentric annuli of outer radii $R_{\mathrm{k}}$, and the dielectric response ($\varepsilon_{\parallel}$) is built outward one shell at a time. In this approach, the homogenized disk of radius $R_{\mathrm{k}}$ becomes the inclusion for the next shell $\varepsilon_{\mathrm{k+1}} = \varepsilon_{\mathrm{Nb}}(r_{\mathrm{k+1}},\omega)$ through the two-dimensional Maxwell–Garnett relation for cylinders with axes perpendicular to the field,

$$\varepsilon^{(k+1)} = \varepsilon_{\mathrm{k+1}}\frac{\varepsilon^{(k)}(1+\varphi_{\mathrm{k}}) + \varepsilon_{\mathrm{k+1}}(1-\varphi_{\mathrm{k}})}{\varepsilon^{(k)}(1-\varphi_{\mathrm{k}}) + \varepsilon_{\mathrm{k+1}}(1+\varphi_{\mathrm{k}})}, \quad \text{where } \varphi_{\mathrm{k}} = (R_{\mathrm{k}}/R_{\mathrm{k+1}})^2 \tag{S13}$$

with $\varepsilon^{(1)} = \varepsilon_{\mathrm{Nb}}(r_1,\omega)$ and $\varepsilon_{\parallel} = \varepsilon^{(N)}$. We note that this recursion is essentially a continuous generalization of the two-phase normal-core model [Eq. (3) of the main text]. Once $\varepsilon_{\perp}$ and $\varepsilon_{\parallel}$ are determined the heat-transfer coefficient between the two homogenized uniaxial half-spaces is evaluated from the uniaxial Fresnel coefficients and the theory of Polder–Van Hove [18].

**Comparison with RCWA.** The graded-cell EMT reproduces the exact-lattice RCWA value of $h$ to within 1% for $\bar{B} \gtrsim 55$ mT, and to within 2.6% at the most dilute lattice computed ($\bar{B} \approx$ 10.3 mT, $a_0 = 482$ nm) as can be seen from representative values given in Table S5. For comparison, we note that the normal-core model of Eq. (3) overestimates $h$ by up to a factor of 2.5 over the same range (Fig. 2(a), lower panel). As mentioned in the manuscript the close agreement of the graded-cell EMT holds because the vortex spacing lies below the near-field scale $2\pi d$.

*TABLE S5. Computed heat-transfer coefficient h ($W\,m^{-2}\,K^{-1}$) from the exact-lattice RCWA and the graded-cell EMT at d = 100 nm and T = 4.2 K.*

| $\bar{B}$ (mT) | $a_0$ (nm) | h (RCWA) | h (graded) | deviation |
|---|---|---|---|---|
| 10.3 | 482 | $8.119\times10^{-3}$ | $7.911\times10^{-3}$ | −2.56% |
| 25 | 307 | $1.135\times10^{-2}$ | $1.1078\times10^{-2}$ | −2.43% |
| 133 | 134 | $4.370\times10^{-2}$ | $4.360\times10^{-2}$ | −0.22% |
| 493 | 69 | $3.357\times10^{-1}$ | $3.3533\times10^{-1}$ | −0.13% |
| 698 | 58 | $8.901\times10^{-1}$ | $8.887\times10^{-1}$ | -0.16% |

**Spectral comparison.** In Figure S1 we present a comparison of the spectral heat flux $h(\omega)$ computed by RCWA and by the graded-cell EMT at five mean inductions $\bar{B}$, together with their ratio. Across the thermal peak, which carries most of the flux, the two agree to about 1%, the EMT lying slightly below RCWA. The EMT is found to overestimate the flux in two regions: 1) at low frequencies, $\hbar\omega \lesssim 0.3$ meV, where the skin depth grows and the near field probes wavevectors approaching the lattice scale, and 2) near the gap edge $2\Delta_0$ for dilute fields, where the local conductivity varies most sharply across the vortex cell. Both regions carry a small fraction of $h$, so the integrated heat transfer agrees to within 0.1–2.6%, the poorer agreement occurring at the most dilute lattice. The graded-cell EMT thus reproduces not only the integrated heat transfer but its spectral distribution wherever the flux is significant.

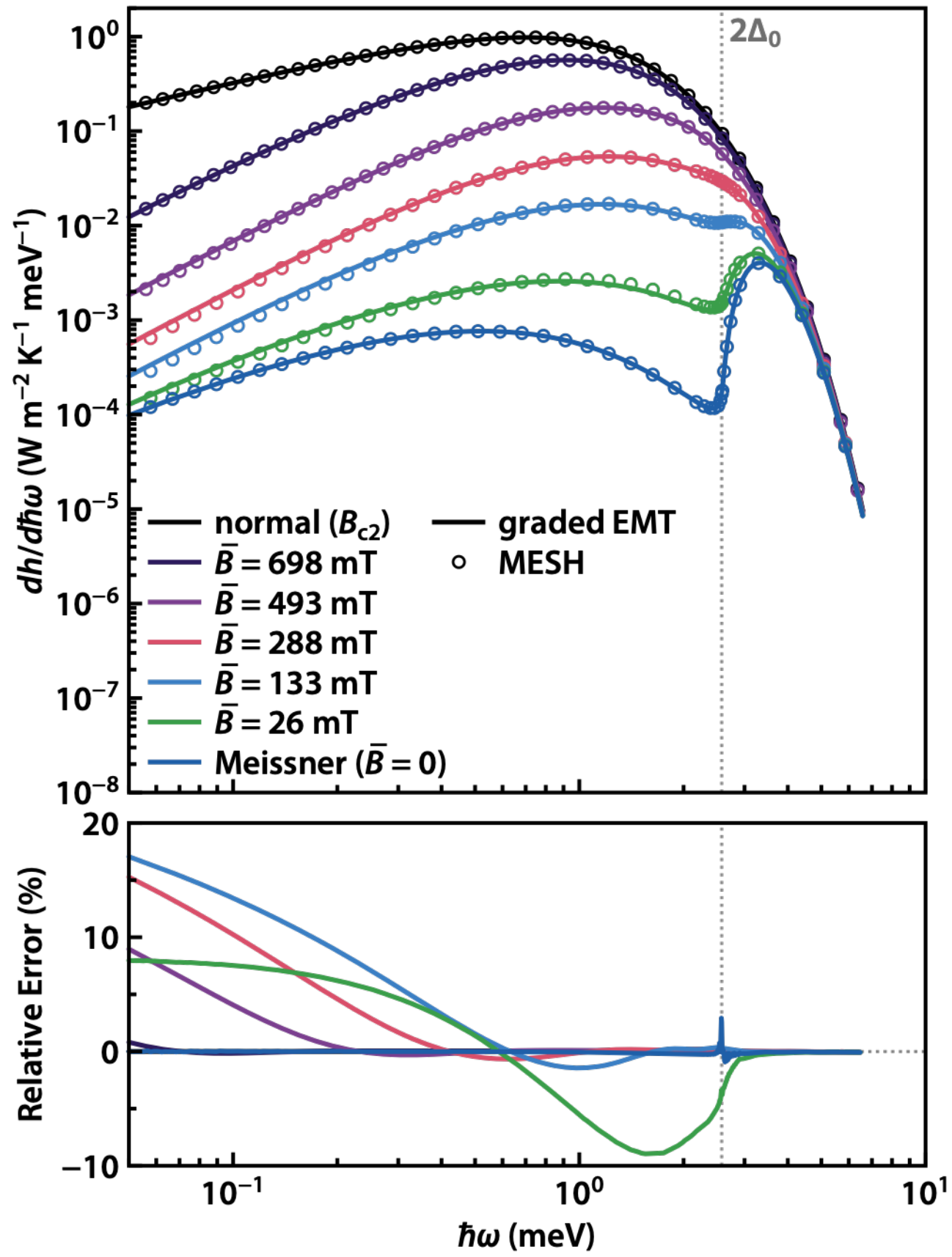

*FIG. S1. Spectral heat flux h(ω) computed by RCWA (circles) and by the graded-cell EMT (lines) at T = 4.2 K and d = 100 nm, for five mean inductions $\bar{B}$. Lower panel shows the relative percent error of EMT with respect to the RCWA spectrum (100× ($h_{graded\text{-}cell\ EMT}$ - $h_{RCWA}$)/$h_{RCWA}$). The two agree to about 1% across the thermal peak, with the EMT lying slightly below RCWA. The EMT overestimates the flux at low frequencies. However, these regions carry a very small fraction of the total flux, so the integrated h agrees to within 0.1–2.6% of the RCWA.*

**The normal-core effective medium theory and the origin of its overestimate of *h*:** The normal-core EMT described in the manuscript models each vortex as a disk of radius $\sqrt{2}\xi$ in which the superconductor is assumed to be fully normal ($f = 0$), and the matrix outside is assumed to be uniform and featuring a pair correlation gap $f(R)\Delta_0$. The pair-breaking scattering rate in the superconducting region is set by the order parameter $f(R)$ at the cell boundary. The combined effect of these two distinct media is captured by the Maxwell–Garnett rule described in the main manuscript. As shown in Fig. 2(a), the normal-core EMT agrees with the RCWA and graded-cell

EMT in the Meissner and normal states but overestimates h in the mixed state by up to a factor of 2.5.

The origin of this overestimation can be understood by noting that the model departs from the exact profile in two ways: 1) it treats the disk of radius $\sqrt{2}\xi$ as fully normal, and 2) it assigns the single cell-boundary gap $f(R)\Delta_0$ to the entire matrix. To identify the impact of these two assumptions we constructed two intermediate effective media. In the first, we assumed that the core retains the true profile $f(r)$ inside $\sqrt{2}\xi$ while the matrix is held uniform at $f(R)$ whereas in the second, we assume that the core is fully normal but the matrix is described by the true $f(r)$.

In Table S6 we present results from the above described modified normal-core models each expressed as a ratio to the graded-cell EMT result. We find that the second variant of the normal-core EMT reproduces the normal-core overestimate almost entirely (2.58 against 2.48 at 395 mT, and to within 1% at 553 and 711 mT), whereas the first lies at or below the graded result at every field. Therefore, the deficiency of the normal-core EMT lies largely in imposing a sharp core of fixed radius, which the graded-cell EMT removes by using the full radial profile of the order parameter.

*TABLE S6. Ratio of the result of each modified effective-medium theory to that of the graded-cell EMT at d = 100 nm, T = 4.2 K.*

| **$\bar{B}$ (mT)** | ***f*(*R*)** | **Normal-core EMT** | **Core graded, matrix uniform** | **Core normal, matrix graded** |
|---|---|---|---|---|
| 40 | 0.998 | 0.88 | 0.66 | 1.37 |
| 79 | 0.990 | 1.03 | 0.60 | 1.65 |
| 158 | 0.953 | 1.53 | 0.70 | 2.08 |
| 276 | 0.872 | 2.19 | 0.86 | 2.47 |
| 395 | 0.770 | 2.48 | 0.95 | 2.58 |
| 553 | 0.599 | 2.22 | 0.99 | 2.23 |
| 711 | 0.347 | 1.47 | 1.00 | 1.47 |

**S6. Computed transmission function for *d* = 100 nm, *T* = 4.2 K**

As stated in the manuscript, past work on NFRHT in metasurfaces has shown that changing the filling fraction of a lattice of fixed periodicity can enhance the transfer through the formation of surface modes [19,20]. The signature of such surface modes is a bright branch in the transmission $\mathcal{T}(\omega, k_{\parallel})$, where $k_{\parallel}$ is the in-plane wavevector of the vacuum plane wave that carries the transfer. To test whether any such mode arises in the vortex metasurface, in which the field changes the periodicity and the filling fraction together, we computed $\mathcal{T}$ at $d$ = 100 nm for four states: the Meissner state, mixed states at $\bar{B}$ = 133 and 493 mT, and the normal state (Fig. S2).

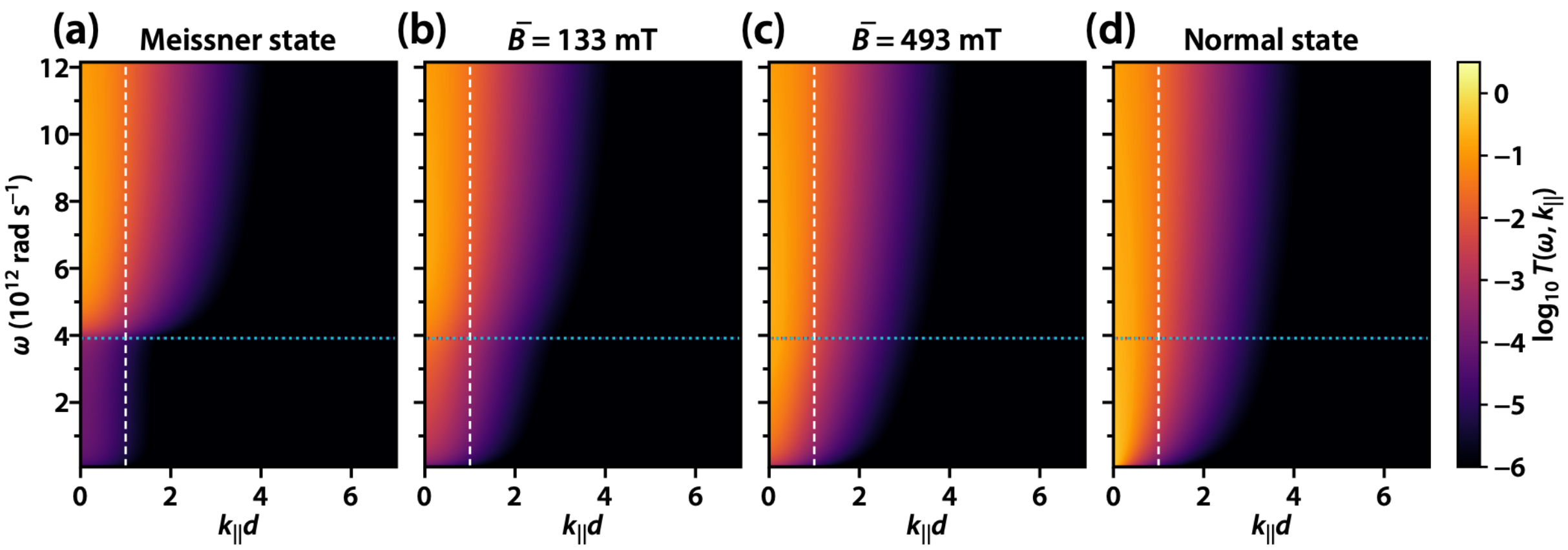

*FIG. S2. Transmission $\mathcal{T}(\omega, k_{/\!/})$ at d = 100 nm and T = 4.2 K, summed over polarizations, for (a) the Meissner state, (b) a dilute lattice ($\bar{B}$ = 133 mT, $a_0$ = 134 nm), (c) a dense lattice ($\bar{B}$ = 493 mT, $a_0$ = 70 nm) and (d) the normal state. All four plots are computed from the graded-cell effective medium theory. The exact-lattice calculation gives the same heat-transfer coefficient to within 1% and the same spectral flux to within a few percent. White dashed line shows the near-field window boundary ($k_{/\!/} = 1/d$). The cyan dotted line marks the superconducting gap $2\Delta_0$ = 2.6 meV.*

The transmission maps are computed from the graded-cell EMT, so that $\mathcal{T}$ is the specular transmission of a homogeneous uniaxial medium and all four panels share the same axes. We note that the RCWA calculation gives a heat-transfer coefficient that is within 1% of that of graded-cell EMT. Similarly, the same spectral flux $h(\omega)$ also agrees within a few percent at every frequency (see Sec. S5), justifying the use of the transmission plots from EMT instead of RCWA. We find that no bright branch appears in any panel. Instead, all four show that NFRHT has contribution from the same broad range of wavevectors with 80% of the heat transfer carried by channels with $k_{\parallel}$ *between* $\frac{0.2}{d}$ *and* $\frac{1}{d}$. We also note that NFRHT is dominated by s-polarized (TE) channels, with the p-polarized (TM) contribution being negligibly small (<$10^{-3}$). The transmission plots show that as the applied field increases, the magnitude of $\mathcal{T}$ increases at sub-gap frequencies. However, above the superconducting gap the four states converge and for $\hbar\omega = 4$ meV they transmit within a factor of two of one another. This convergence in the transmission occurs since photons of these energies break Cooper pairs and the superconductor absorbs comparably to a normal metal. The enhancement produced by the applied field is therefore dissipative and broadband rather than resonant and is in contrast to that observed in past work on metasurfaces [19,20].

**S7. Discussion on the origin of the maximum in the switching contrast as a function of gap size**

For the Meissner and normal states the transmission in Eq. (2) of the manuscript separates into *s*- and *p*-polarized parts, $\mathcal{T} = \mathcal{T}_s + \mathcal{T}_p$, with

$$\mathcal{T}_\alpha(\omega, k_\parallel) = \frac{4\,(Im\,r_\alpha)^2\,e^{-2\gamma d}}{|1 - r_\alpha^2\,e^{-2\gamma d}|^2}, \;\; \gamma = (k_\parallel^2 - k_0^2)^{1/2}, \quad \text{(S14)}$$

for evanescent waves, where $\alpha = s, p$ labels the polarization and $r_s$ and $r_p$ are the Fresnel reflection coefficients of a single surface computed from $\varepsilon_{Nb}(\omega)$ and $k_0 = \omega/c$. For propagating waves, $k_\parallel < k_0$, the corresponding expression is

$$\mathcal{T}_\alpha(\omega, k_\parallel) = \frac{(1 - |r_\alpha|^2)^2}{\left|1 - r_\alpha{}^2\,e^{2ik_z d}\right|^2}, \;\; k_z = \left(k_0{}^2 - k_\parallel{}^2\right)^{1/2}, \quad \text{(S15)}$$

which carries no exponential attenuation across the gap and vanishes for a perfectly reflecting surface, for which $|r_\alpha| = 1$. Figure S3(a),(b) shows the *s* and *p* contributions to *h* as a function of *d* in the two states. In both, the *s*-polarized flux dominates at small gaps. It saturates once *d* falls below the length over which the field penetrates the surface — the skin depth $\delta$ in the normal state and the penetration depth $\lambda$ in the Meissner state — and decays for larger gaps, with a slope that steepens with *d* and approaches $d^{-4}$ only at the largest gaps shown. The *s* channel alone would thus give a contrast that one would expect to rise monotonically with *d*.

The *p*-polarized flux is much weaker but decays only as $d^{-1}$ over the whole range, because it is carried by wavevectors close to the light line, for which the attenuation across the gap is weak. It contributes only at frequencies where the surface absorbs, since a lossless surface reflects these waves completely. In the normal state the *s* channel is strong and the *p* channel overtakes it only at $d \approx 36$ µm. In the Meissner state the $s$ channel is strongly suppressed by the small sub-gap absorption Im $\varepsilon_{Nb}$ (see Fig. 1(c) of the main manuscript), whereas the $p$ channel is carried mainly by photons above $2\Delta_0$, which the superconductor absorbs comparably to a normal metal, so it is only modestly suppressed. The crossing therefore occurs at a much smaller gap, $d \approx 6$ µm. Between the two crossings the Meissner-state flux decays as $d^{-1}$ while the normal-state flux continues to fall steeply, so the switching contrast decreases, as seen in Fig. 3(b) of the main text. The maximum of the contrast, about 1500 at $d \approx 3.5$ µm, occurs before the Meissner-state crossing as at that gap the slowly decaying $p$ channel is already a significant fraction (about 20%) of the Meissner flux, enough to halt the steep decline of the total, while the normal-state flux is still dominated by its steeply falling $s$ channel.

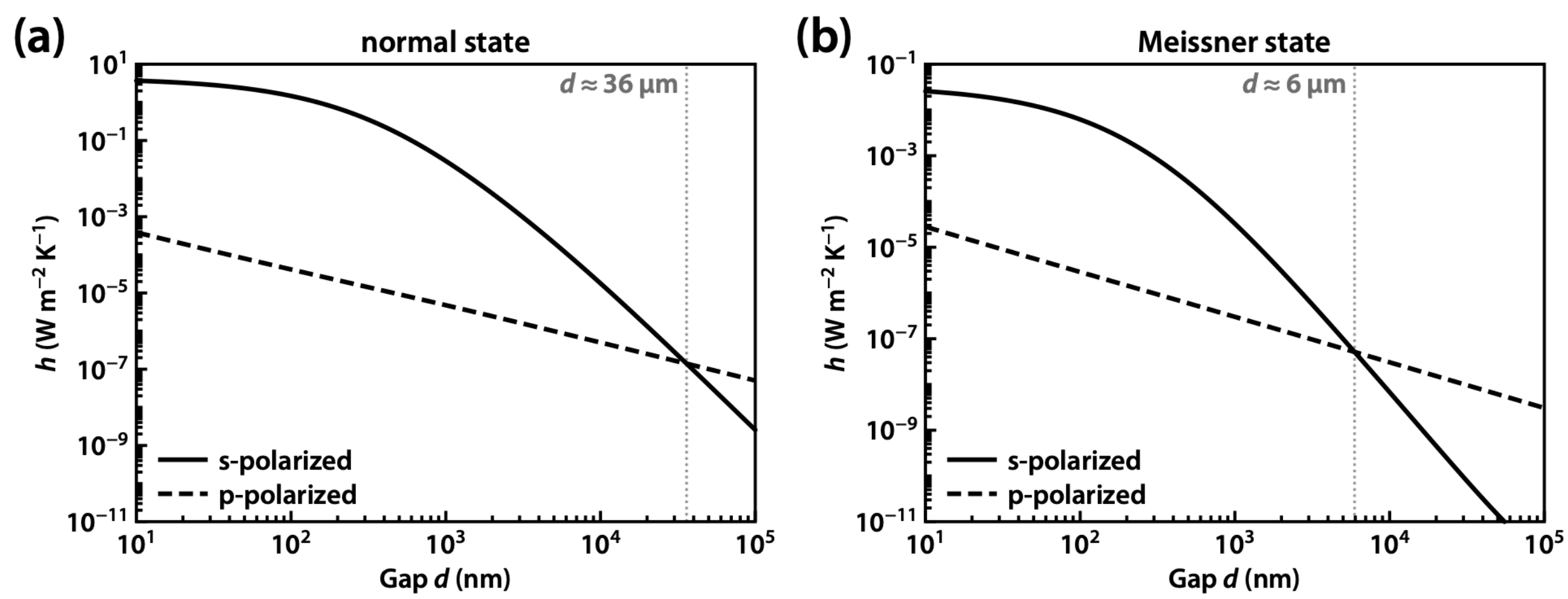

*FIG. S3. Contributions of s- and p-polarized waves to NFRHT at T = 4.2 K in (a) the normal state and (b) the Meissner state. The s-polarized (solid) and p-polarized (dashed) contributions are shown as a function of the gap size d. The dotted lines mark the gap size at which the relative contribution of s- and p-polarized waves flips (d ≈ 36 µm (normal) and d ≈ 6 µm (Meissner)).*